# Calculate Center-of-Inertia Frequency and System RoCoF Using PMU Data

Shutang You, Hongyu Li, Shengyuan Liu, Kaiqi Sun, Weikang Wang, Wei Qiu
*Department of Electrical Engineering and Computer Science*
*The University of Tennessee, Knoxville*
Knoxville, TN, USA
syou3@utk.edu

Yilu Liu
*The University of Tennessee, Knoxville and Oak Ridge National Laboratory*
Knoxville and Oak Ridge, TN, USA
liu@utk.edu

***Abstract*— The power system frequency is important for the system overall stability. However, there does not exist a single measurement point of the system frequency due to the distributed nature of the system inertia and the small inconsistency of different generator rotors' electrical speeds in one synchronized system. This paper proposed a new approach to calculate the system center-of-inertia (COI) frequency and the rate-of-change-of-frequency (RoCoF) more accurately using PMU data at multiple locations. The COI frequency and the RoCoF value were further used to assist fast estimation of the imbalance MW amount of a frequency event. Test results using actual measurements in the U.S. Eastern Interconnection system validated the effectiveness of the proposed method.**

***Keywords—COI, PMU, RoCoF, synchrophasor, power system inertia***

## I. INTRODUCTION

As one of the most critical indices, the power system frequency directly reflects the real-time balance condition between system generation and load. Due to its importance, the system frequency is also a key attribute monitored by wide-area measurement systems (WAMSs). Because of the wide geographical distribution of power system facilities and the existence of electromechanical oscillations and wave propagation, frequencies at different locations are usually different and thus difficult to be fused to obtain the system-level frequency [1, 2]. This is especially challenging during large disturbances, despite the fact that obtaining the accurate frequency of the system during large disturbances is critical. Moreover, the limited number of synchrophasor sensors and limited observability of system at the synchrophasor time-resolution also make it difficult to obtain the system-level frequency [3, 4].

To overcome this challenge, the center of inertia (COI) frequency, which is defined as the frequency using electric rotor speeds weighted by rotor inertia values, has been proposed to represent the true system frequency. However, it is almost impossible to obtain real-time speed information from all inertia-contributing generator rotors in actual power systems. Fig. 1 shows an example of the measured frequency (in Hz) of a generation trip event in the U.S. East Interconnection (EI) grid. The existence of oscillations and electromechanical wave propagation can be seen from the zoomed-in frequency plot on the top right of Fig. 1. Actually, the discrepancy of frequency measurements at different locations can be observed in almost all events in multi-machine power systems, especially for large-scale systems. Therefore, it is difficult to obtain the COI frequency and the rate-of-change-of-frequency (RoCoF) value for the whole system [5, 6], although the system COI frequency and RoCoF are found to be very useful for many power system monitoring and control functions, which include under-frequency load shedding [7], frequency response analysis [8], frequency control [9], event magnitude estimation location estimation [10, 11], transient stability improvement and fault-ride-through control [12], and resource size and location selection [13], etc.

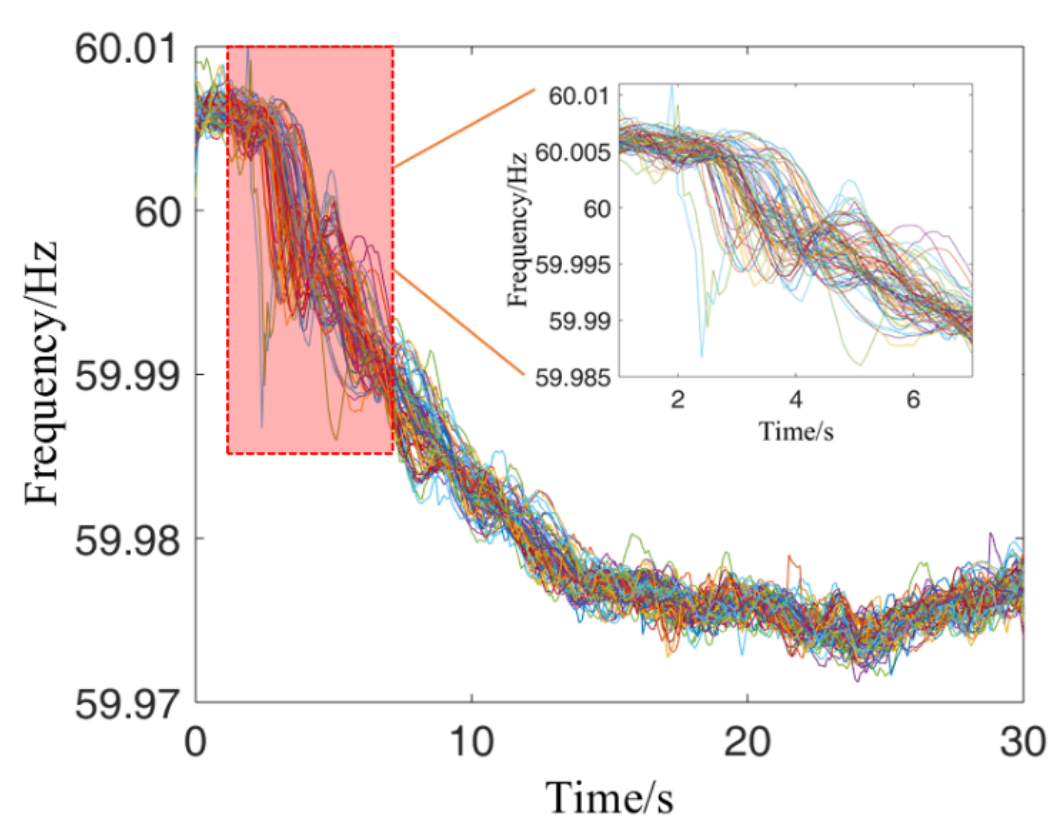

Fig. 1. Frequency measurements of a generation trip event in the U.S. Eastern Interconnection grid recorded by FNETGridEye on 01/18/2018 [14]

Some work has noticed the importance and challenges of obtaining the accurate COI frequency and RoCoF, and some has made some remarkable progress. Ref. [13] calculated the coherence degree of all buses in the system and identified buses that have low contribution to the COI frequency. Ref. [5] developed a method to obtain the COI frequency using bus frequency, transmission system parameters, and generator parameters. Ref. [15] used a model decoupling method and Kalman filter to estimate the generator rotor speed and inertia, which were then used to derive the system COI frequency. These studies are significant advancements of COI frequency estimation considering different availability levels of system measurements.

In a deregulated power system, the transmission-level measurement and parameters are usually proprietary and not sharable with other entities. The only available measurement in

This work was supported in part by the Engineering Research Center Program of the National Science Foundation and the Department of Energy under NSF Award Number EEC-1041877 and the CURENT Industry Partnership Program. Funding for this research was partially provided by the NSF Cyber-Physical Systems (CPS) Program under award number 1931975 and the U.S. Department of Energy under Award Number 34231.

such deregulated market structures may be the frequency measurement from sensors deployed at the distribution level, such as the FNET/GridEye system [16]. Obtaining the system COI frequency solely based on these data has significant value for balancing authorities, reliability coordinators, and regulation entities to help understand the system real-time balancing condition. To solve this issue, this paper proposes a method to calculate the COI frequency and RoCoF using synchrophasor data from WAMSs based on distribution-level synchrophasor sensors. The merits of the new approach include: i) it is more robust to the changes of power grid conditions and the deployment (i.e., number and locations) of sensors; ii) it can obtain a more accurate estimation of the COI frequency and the RoCoF value.

## II. Review of COI Frequency and ROCOF Calculation

Theoretically, the COI frequency can be calculated using the machine speed and the inertia value of each committed generation unit in power systems. The COI frequency ($f_{\mathrm{COI}}$) is the average electrical speed of all machines weighted by their inertia values [17].

$$f_{\mathrm{COI}} = \frac{\sum_{i=1}^{N} I_i \cdot f_i}{\sum_{i=1}^{N} I_i} = \frac{1}{I_{sys}} \sum_{i=1}^{N} I_i \cdot f_i \tag{1}$$

where $I_i$ and $f_i$ are the inertia time constant and the real-time speed of the $i^{\mathrm{th}}$ machine; respectively. $I_{sys}$ is the system total inertia and equals $\sum_{i=1}^{N} I_i$.

ROCOF is defined as the first-order derivative of the frequency. IEEE Std. C.37.118.1 specifies that PMUs need to provide RoCoF data at a reporting rate of 10 data per second with the steady-state error limited to 0.01Hz/s [18]. For reliability regulation purposes, the calculation method of RoCoF proposed by North American Electric Reliability Corporation (NERC) is to calculate the frequency change in the first 0.5s after a frequency disturbance [19]:

$$\text{RoCoF (or } df/dt) = (f_0 - f_{0.5})/0.5 \tag{2}$$

where $f_0$ and $f_{0.5}$ are the frequencies at the event start time and 0.5 second after the event occurrence, respectively. This RoCoF calculation method is practical and robust when using the frequency data at one location.

However, as mentioned before, the real-time speed values of all machines are not always available in practice due to insufficient measurement coverage in large power grids. This makes it difficult to obtain the COI frequency. The estimation of RoCoF is also difficult to be applied in large power systems since RoCoF values appear different for different locations [20]. In addition, it is nontrivial to select the time duration (for example, the tentative 0.5s suggested by NERC) for RoCoF calculation in specific power systems considering the variations in system conditions and event magnitudes.

## III. Center of Inertia (COI) Frequency and RoCoF Calculation

### A. *Measurement System Introduction — FNET/GridEye*

To study the COI and RoCoF calculation problem, the data used in this study is from the wide-area frequency monitoring network called FNET/GridEye. This is a synchrophasor measurement system that collects real-time frequency, voltage magnitude, and phasor angle data using a sensor called Frequency Disturbance Recorders (FDRs, as shown in Fig. 2). The measured frequency and voltage data are time-stamped using GPS signals and then transmitted to the data server located in University of Tennessee and Oak Ridge National Laboratory. As of 2020, more than 300 FDRs have been deployed into the power grids around the world. The deployment locations of FDRs in the North American power grids are shown in Fig. 3. The world-wide coverage of FDRs is shown in Fig. 4. The frequency measurement data are shown in real-time on the FNET/GridEye Public website (http://fnetpublic.utk.edu/tabledisplay.html). Based on FNET/GridEye, many real-time and off-line applications have been developed, such as real-time power grids frequency and voltage monitoring [16], frequency event detection and reporting, oscillation detection and analysis [21], islanding detection [22], inertia estimation [4, 23], model validation [24], responsive load control [25], and cyber-attack detection [10, 26], etc. These application provide a variety of functions for power grid situational awareness and reliability improvement.

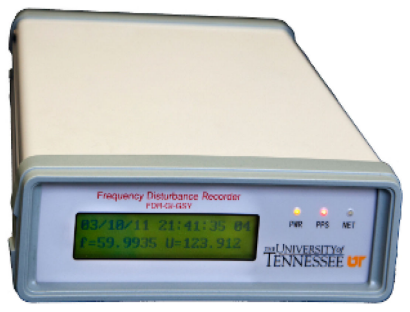

Fig. 2. Frequency Disturbance Recorder (FDR)

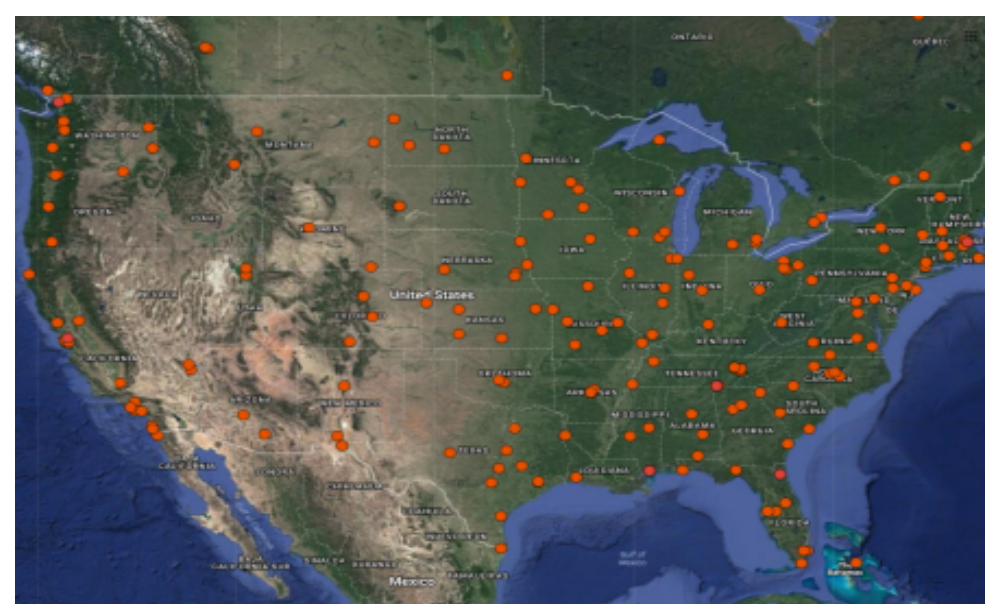

Fig. 3. Deployment locations of FDRs in North American power grids

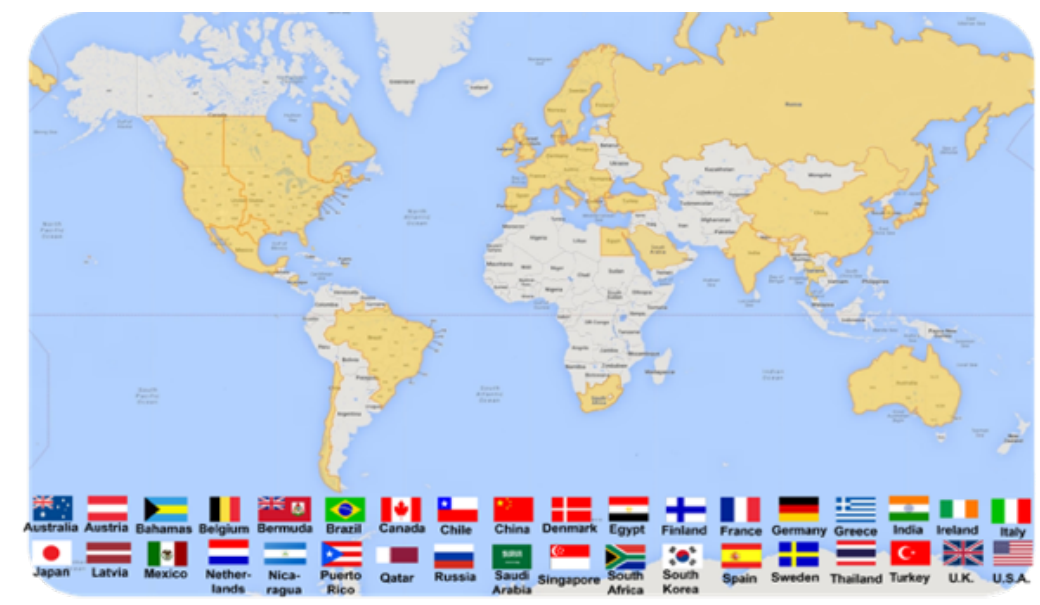

Fig. 4. Countries with FDRs deployed in power grids

Fig. 5 shows an example of the event report provided by FNET/GridEye. This event report gives information of a generation trip event that occurred on August 6, 2018 in the U.S. EI system. The report includes frequency plots at different locations, and more importantly, the event location and magnitude estimation. In this generation trip event, the frequency measurement collected near the generation trip location in the MISO area had a large dip down to around 59.96Hz, and quickly bounced back due to the support from the rest of the grid. Meanwhile, the system frequency declined to 59.98Hz before frequency regulation facilities (i.e. Automatic Generation Controls) brought the frequency back to 60Hz gradually. This event report provided important information to operators to better understand the real-time system situation.

Basic Event Information

| Event Date | Event Time | Event Type | Estimated Amount |
| --- | --- | --- | --- |
| 2018-08-06 | 16:12:47 UTC | Generation Trip | 690 MW |
| Point A | Point B | Point C | Point C Prime |
| 60.009 Hz | 59.9868 Hz | 59.9809 Hz | N/A Hz |
| MOD-027-1 Event | Inter Connection | Estimated Reliability Coordinator | ROCOF |
| NO | EI | MRO | N/A |
| Estimated Event Location | | Additional Location Information | |
| (47.2611, -93.6528) | | near Clay Boswell power plant (MRO) in (Cohasset,MN,55721). | |
| Unit Detection Order (the first 6 units) | | | |
| UsMnGre790, UsMnMisostpaul832, UsMnDodgeCenter877, UsNdMdudickinson729, UsNdMduwilliston726, UsMTMduglendive730 | | | |

Fig. 5. An example of the event report in FNET/GridEye

## B. *COI Frequency Calculation*

Having recorded thousands of events like the one shown in Fig. 5. FNET/GridEye serves as a rich resource of synchrophasor data in calculating the system COI frequency. Due to the existence of governor deadband and time delay of governor response, the response of governors is insignificant during the initial period after a generation trip event (in around 1s after a frequency event, depending on system inertia and event magnitudes). The system load has not responded much to this initial small frequency deviation either since the system frequency change is very small within one second in common frequency events. Therefore, the system frequency drop is assumed to be only caused by the MW generation loss of a generation unit trip.

Inspired by the definition of COI frequency, the weighted frequency measurement is used to calculate the system-wide frequency. Let $\{f_1[t], f_2[t], \ldots, f_N[t]\}$ represent the frequency measurement at different locations and $t = T_1, T_2, \ldots, T_K$ represent the timestamps of the frequency measurement after the frequency event occurrence. (The generation trip event start time $T_0$ and event-start frequency $F_0$ can be obtained accurately using Delaunay triangulation and bicubic 2D interpolation based on our previous work in [10].) Fig. 6 provides an illustration graph of the COI frequency position (represented by the solid blue line) after the generation trip event shown in Fig. 1.

The inertia weights (which are dimensionless values) of frequency measurements are denoted as $\{x_1, x_2, \ldots, x_N\}$. Since the COI frequency will have a constant RoCoF, the above variables have the following equations.

$$\begin{cases} f_1[T_1] \cdot x_1 + f_2[T_1] \cdot x_2 + f_3[T_1] \cdot x_3 + \cdots + f_N[T_1] \cdot x_N = F_0 + \Delta F \\ f_1[T_2] \cdot x_1 + f_2[T_2] \cdot x_2 + f_3[T_2] \cdot x_3 + \cdots + f_N[T_2] \cdot x_N = F_0 + 2 \cdot \Delta F \\ \vdots \\ f_1[T_K] \cdot x_1 + f_2[T_K] \cdot x_2 + f_3[T_K] \cdot x_3 + \cdots + f_N[T_K] \cdot x_N = F_0 + K \cdot \Delta F \end{cases} \quad (3)$$

where $\Delta F$ is the system frequency deviation between adjacent data points and $F_0$ is the starting frequency of the event.

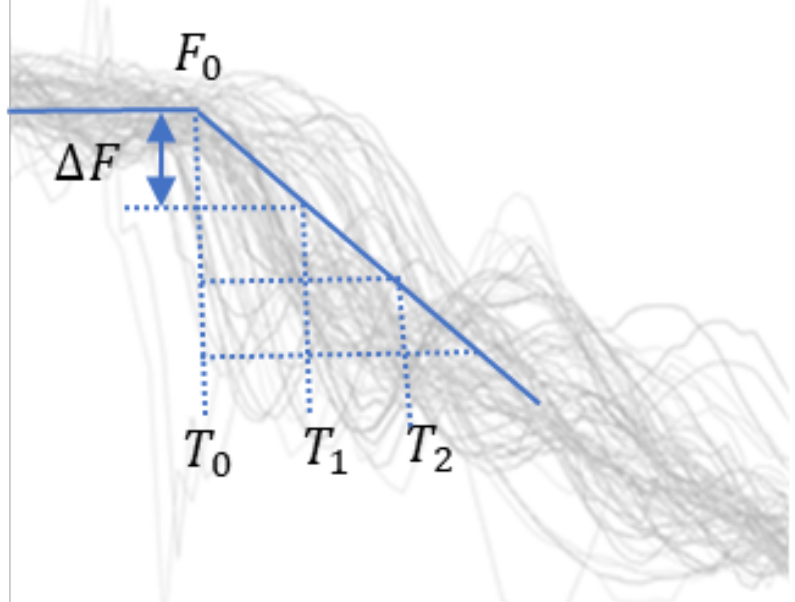

Fig. 6. The COI frequency after a generation trip (the background frequency plot is from the zoomed-in part in Fig. 1)

The sum of all weights $\{x_1, x_2, \ldots, x_N\}$ of frequency measurement equals to one and each weight $(x_n)$ of frequency measurement varies around $1/N$, where $N$ is the total number of frequency measurement sensors. The initial guess of weighting values is based on the average weighting strategy, in which each frequency sensor measures the frequency of a same percentage of system mechanical rotors.

$$\begin{cases} \sum\{x_1, x_2, x_3, \ldots, x_N\} = 1 \\ x_i \approx 1/N, \quad for\ i = 1,2, \ldots, N \end{cases} \quad (4)$$

If the frequency measurement has a high-enough time resolution, (i.e., $K > N + 1$, which is easily achievable due to the availability of high reporting rate synchrophasor measurement data), then, (3) and (4) will constitute an overdetermined system with relaxed constraints on weights. Due to the heterogeneity of generation, transmission, and sensor distribution, $\{x_1, x_2, \ldots, x_N\}$ are unequal and constraint (4) should be relaxed. This relaxation is adjusted by a weighting factor $\omega$, which can be determined based on system-specific experience to strike a balance between the linearity of the initial frequency drop and deviation of weights from $1/N$. Then the solution of this system can be found using the ordinary least square method:

$$\min \|\boldsymbol{A}\boldsymbol{x}_{\boldsymbol{E}} - \boldsymbol{b}\| \quad (5)$$

where

$$\boldsymbol{A} = \begin{bmatrix} f_1[T_1] & f_2[T_1] & \cdots & f_N[T_1] & -1 & -1 \\ f_1[T_2] & f_2[T_2] & \cdots & f_N[T_2] & -2 & -1 \\ \vdots & \vdots & \vdots & \vdots & \vdots & \\ f_1[T_K] & f_2[T_K] & & f_N[T_K] & -K & -1 \\ \omega & \omega & & \omega & 0 & 0 \\ \omega & 0 & \cdots & 0 & 0 & 0 \\ 0 & \omega & & 0 & 0 & 0 \\ \vdots & \vdots & & \vdots & \vdots & \vdots \\ 0 & 0 & & \omega & 0 & 0 \end{bmatrix} \quad (6)$$

$$\boldsymbol{x}_{\boldsymbol{E}} = [x_1, x_2, \cdots, x_N, \Delta F, F_0]^T \quad (7)$$

$$\boldsymbol{b} = [0,0, \cdots ,0, \omega, \omega/N, \ldots, \omega/N]^T \quad (8)$$

The solution of this overdetermined system is

$$\boldsymbol{x}_{\boldsymbol{E}} = (\boldsymbol{A}^{\boldsymbol{T}}\boldsymbol{A})^{-1}\boldsymbol{A}^{\boldsymbol{T}}\boldsymbol{b} \quad (9)$$

After $\boldsymbol{x}$ is determined ( $\boldsymbol{x} = [x_1, x_2, \cdots, x_N]^{\boldsymbol{T}}$ , which are the weights of different PMU channels), the COI frequency can be calculated by weighting the frequency values as

$$f_{\text{COI}} = (\boldsymbol{f} \times \boldsymbol{x}) / \textstyle\sum_{n=1,\ldots,N} x_n \quad (10)$$

where $\boldsymbol{f}$ is the vector of frequency measurement from PMUs. The RoCoF value is the average frequency change in one second and can be denoted as

$$\text{RoCoF} = \Delta F / \Delta T \qquad (11)$$

where $\Delta T$ is the time duration between two consecutive frequency measurements.

As generation units' inertia time constant, capacities, and their commitment statuses are known parameters for each balancing authority, the system inertia is becoming an accessible and monitored index in control rooms of many power grids [27, 28]. The total inertia from generation can be calculated as

$$I = \sum H_i \cdot Cap_i \qquad (12)$$

Using RoCoF and system inertia information, event magnitude $P_{Event}$ can be estimated right after the event as

$$P_{Event} = P_{Gen} - P_{Load} = \frac{2I}{f_N} RoCoF \qquad (13)$$

where $P_{Gen}$ and $P_{Load}$ are the total power generation and load, respectively. $f_N$ is the nominal frequency (e.g., 60 Hz in the U.S.). $I$ is the system total inertia. The estimated event magnitude information can be used in fast frequency control since the event magnitude information can be obtained right after the event occurrence and before the arrival of the frequency nadir [29].

## IV. CASE STUDY

The effectiveness of the proposed method is assessed using FNET/GridEye measurements [16] with confirmed event magnitude and inertia data in the U.S. Eastern Interconnection grid. The proposed method is compared with the median frequency method, which takes the median value of all frequency measurements as the system frequency. The relaxation parameter $\omega$ in the proposed method is set to 30 for the EI system based on monitoring experience.

Two generation trip events at different locations: Florida (location A in Fig. 7, event shown in Fig. 8) and Kentucky (location B in Fig. 7, event shown in Fig. 9), are used to compare the RoCoF calculation methods. Their frequency measurements are shown in Fig. 8 and Fig. 9, respectively. Thick blue lines in Fig. 8 and Fig. 9 are the COI frequencies obtained using the proposed method. Blue dash lines represent the RoCoF obtained by the proposed method at the beginning of the event occurrence. Thick magenta lines represent the RoCoF obtained by the median frequency method. It can be seen that the generation trip that happened at the grid edge (i.e., Florida) has large oscillations, while the generation trip event happened in the middle of the grid (i.e., Kentucky) has much smaller oscillations. The oscillation in Fig. 8 induces a large difference between the median frequency value and the COI frequency obtained by the proposed method. In contrast, the discrepancy is less visible when the oscillation is insignificant.

With system inertia information, the proposed COI and RoCoF value calculation method can be used to estimate the system event magnitude and reversely verify the accuracy of the COI frequency and RoCoF. The proposed COI frequency and RoCoF calculation method, the frequency deviation method [8], and the median frequency-based method are compared using 86 events of the U.S. EI system in 2016. Statistic results show that the proposed method has 48% and 35% less error, compared with the other two methods respectively.

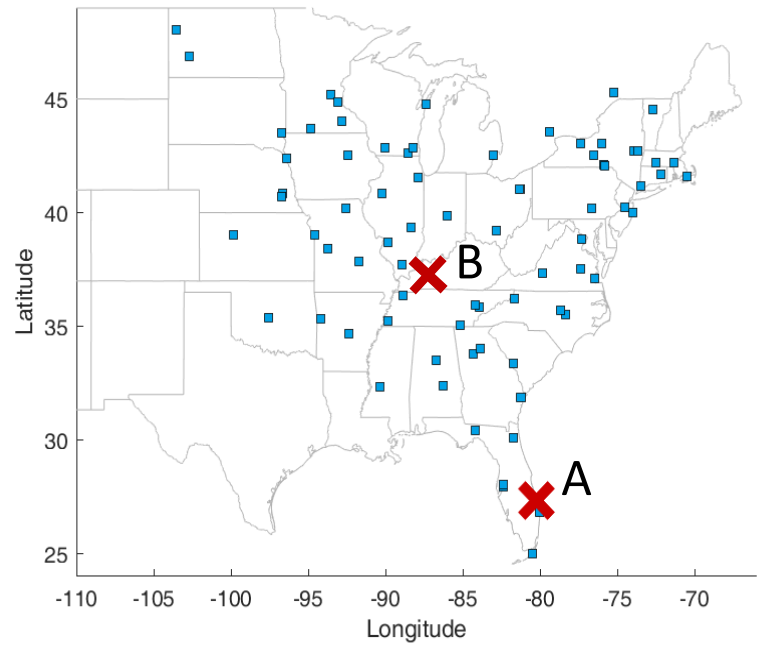

Fig. 7. The locations of two events in the EI (Event A is in Florida; Event B is in Kentucky; blue squares are sensor locations)

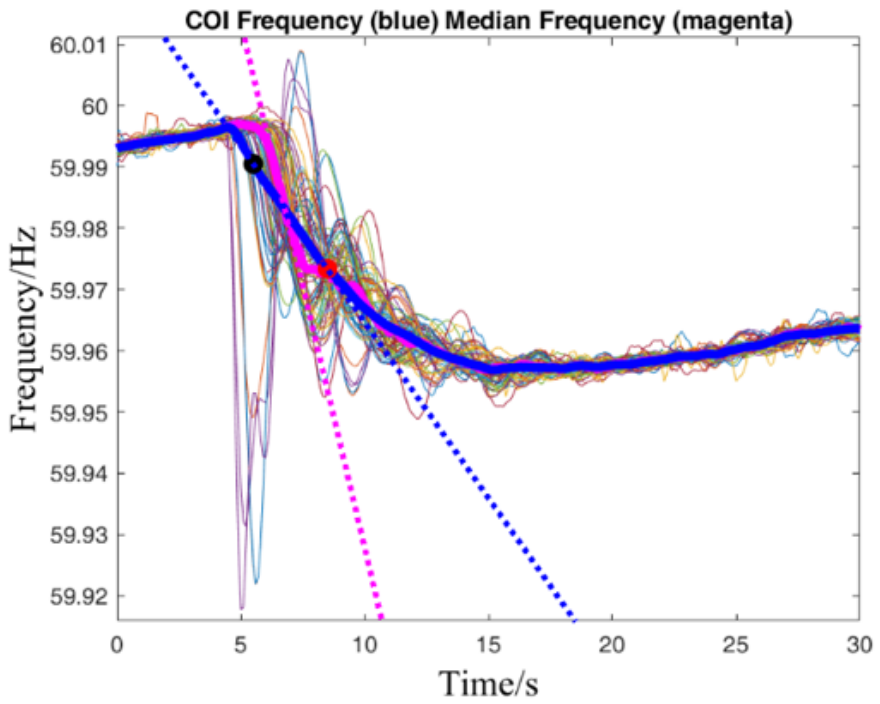

Fig. 8. Generation trip event with large oscillation (Event A happened in Florida, time: 2017-10-26 06:12:55 UTC, 1,010 MW generation loss)

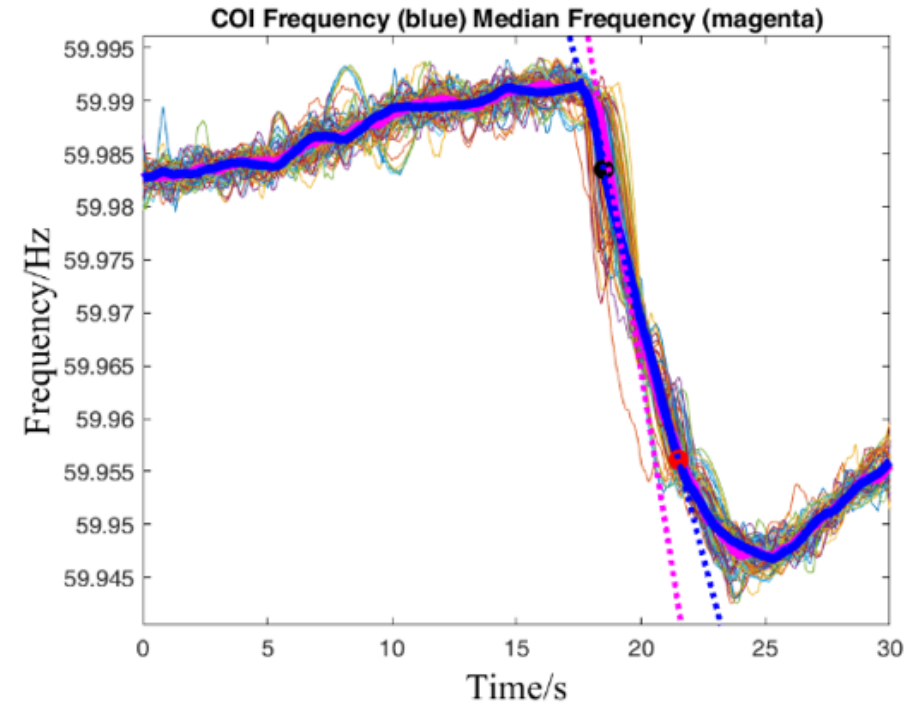

Fig. 9. Generation trip event without large oscillation (Event B happened in Kentucky, time: 2018-04-13 03:51:09 UTC, 900 MW generation loss)

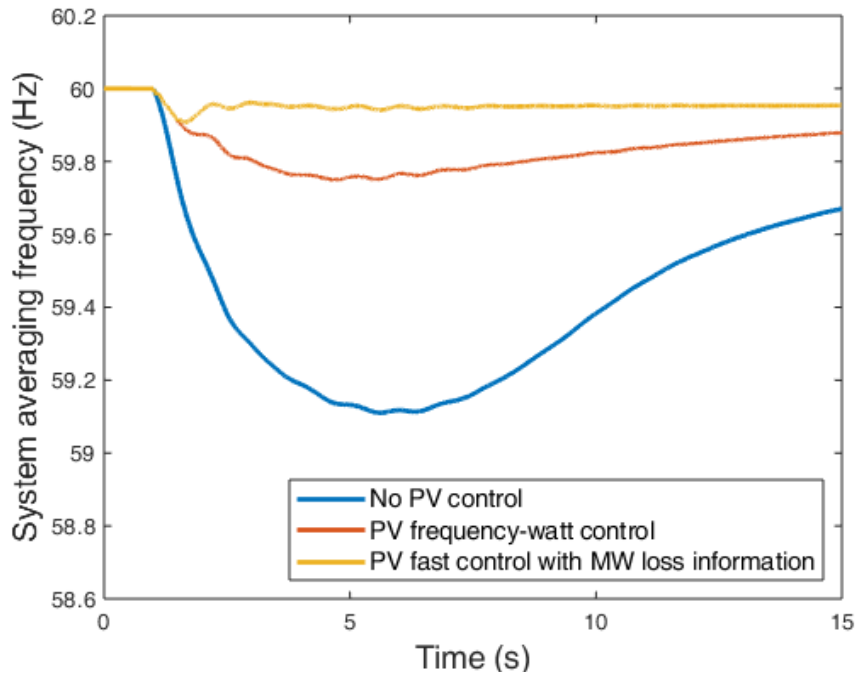

Fig. 10. System frequency response using MW loss information and PV fast frequency control

Once the event magnitude estimation result is available, inverter-based (such as PV and energy storage) fast frequency control can be applied to fast regulate frequency. Fig. 10 shows the system frequency when using MW loss information for PV fast frequency control. It is compared with two other cases: one is the common frequency-watt control [30] without MW loss information and the other is without any inverter-based frequency control. It can be seen that the system frequency response with inverter-based fast frequency control using MW loss information has a higher frequency nadir and better frequency response compared with the common frequency-watt control without MW loss information. This result demonstrated potential of the proposed COI frequency and RoCoF calculation method in improving the system frequency control.

## V. Conclusions

This work proposed a method to calculate the system COI frequency and the RoCoF value using synchrophasor measurements. The method estimates the weight of each frequency measurement in the system COI frequency by solving an over-determined problem. Real-world data from FNET/GridEye were used to validate its effectiveness. Results show the proposed COI and RoCoF calculation method has better performance compared with the median value based method, especially during frequency events with high oscillation magnitudes. The proposed method is also shown to improve event magnitude estimation, which can potentially be used by inverter-based resources for fast frequency control.

## Acknowledgments

We would like to thank colleagues at the University of Tennessee, Knoxville for their help and advices. We also would like to thank our FDR hosts at universities, companies, and individual residents for their participation in data collection.